# A 4641Da polymer of amino acids in Acfer 086 and Allende meteorites

Julie E. M. MCGEOCH[1] and Malcolm W. MCGEOCH[2]

[1] Dept. of Molecular and Cellular Biology, Harvard University, 52 Oxford St., Cambridge MA 02138, USA.
[2] PLEX LLC, 275 Martine St., Suite 100, Fall River, MA 02723, USA

## Abstract

A specific 4641Da amino acid polymer entity is present in two CV3 meteorites Acfer 086 and Allende, together with its breakdown polymer fragments of mass < 2000Da. Previously in Allende and Murchison meteorites we had detected polymer amide of mass < 2000Da using a new method based on Folch extraction of micron meteoritic particles with hydrophilic and hydrophobic phases. Low temperature extraction kept polymer amide intact whereas previous methods had resulted in degradation via boiling the samples. Acfer 086 and Allende polymer amide composition was predominantly glycine with partial content of hydroxy-glycine and alanine. Detection was via matrix-assisted laser desorption time-of-flight mass spectrometry (MALDI/TOF/MS). Amino acid analysis showed that the amino acids present were almost entirely bound in polymers prior to acid hydrolysis. Fragment analysis provided a possible subunit structure for the 4641 species. In the Allende polymer fragment data heavy isotope satellite enhancement confirmed an extraterrestrial origin that could be in the interstellar medium or the protoplanetary disc. The dominance of 4641Da polymer amide suggested a replication or concentration mechanism. Analysis of the infrared modes of a proposed structure showed a peak in collective mode intensity at 20 residue length, giving a possible explanation in terms of optimized energy absorption for the dominance of this size of linear glycine-based structure.

## Introduction

Using a new method of analysis we recently discovered amino acid polymers in two carbonaceous chondrites, Allende and Murchison [1] (henceforth MM2015). In one of these, Allende, we found statistical evidence for polymer comprising glycine, hydroxy-glycine and alanine, bound in varying polymer lengths up to 20 amino acid residues. Previously a very wide range of individual "free" amino acids, most of them non-proteinogenic, had been documented in meteorites, dating back to a paper by Kvenvolden et al. [2] and now the subject of a number of reviews and commentaries [3,4,5,6,7]. As noted in [6] the amino acid extraction protocol was very constant in almost all of this work, depending upon release of amino acids via hot water extraction for 24hours at 100C. In MM2015 the rate of peptide bond hydrolysis in 100C water was assessed to be high enough to have hydrolyzed a fraction of bonds within possible amino acid polymers, especially if the solution was rather strongly alkaline due to meteoritic alkali content. It was therefore possible that at least some of the "free" amino acids observed in prior work had been produced by degradation of polymer amide during the extraction process. Throughout the published data a further fraction of amino acids that had been released upon acid hydrolysis was referred to as a "bound" fraction, but the source of these has not been discovered [3]. It was proposed that the "bound" amino acids are in fact products created in strong acid conditions from precursor molecules [8] but that they could have been released by peptide bond hydrolysis of polymer amide received little discussion. In contrast to prior methods, the method we have used to observe polymer amide, described in detail below, employs grinding out an internal sample of meteorite using a vacuum-brazed diamond burr, followed by room-temperature extraction of the fine particles in a Folch 2-phase solvent mixture [9], then MALDI (matrix-assisted laser desorption of ions) mass spectrometry to give a highly sensitive view of molecular entities with mass-to-charge ratio up to 50,000.

As background we describe the provenance of the CV3 class of carbonaceous chondrite of which two representatives, Allende and Acfer 086, produced the polymer amide of the present report. Carbonaceous chondrites with composition resembling the Vigarano meteorite are designated 'CV' [3] and a designation of level 3 for petrological alteration is the lowest level, implying both low aqueous alteration to the meteorite's material while within its parent body and also low thermal metamorphism. Based mainly upon spectrum similarities in the visible and infrared [10] but also upon the match in density and radar signature [11] it has been tentatively suggested [11] that the common parent body of CV class meteorites could be Asteroid 21 Lutetia, which belongs to the low-albedo "C" group of asteroids. This asteroid has an average diameter of about 100km, sufficiently large to develop a high internal degree of heat via radioactive decays and to support a convective circulation [12]. The paleo-magnetism of Allende may indicate that its parent body was sufficiently large to have an iron core [11]. Although hot at its center for tens of My after formation, the surface temperature would have been <0C throughout its existence, with cool material extending inward for 5 - 10km. A meteorite ejected from its surface would likely contain material that was very little altered from the asteroid's original accretion out of interstellar material within the proto-planetary disc. Indeed, Zolensky et al. [13] describe both Vigarano and Allende CV3 meteorites as sparsely altered, having similar olivine grains both in the matrix and chondrule rims. Early work with Allende failed to detect amino acids [14] but two reports [15,16] have since described a low to moderate abundance. A survey of amino acid abundances across different meteorite types [7] places CV meteorites toward the lower end of the range, with the sum of proteinogenic amino acid concentrations adding up to less than 1500ppb (mass ratio). In summary, Allende is from the least altered class of accreted material, probably never having been exposed (in bulk) to a sustained temperature above 0C. Based on the dates of their chondrules an age progression of chondrite formation may be deduced (from oldest to youngest) CV > CM > CO + OC > CR [17] and therefore the CV3 class promises to hold a record of well-preserved chemistry from the very formation of the solar system, if not from pre-solar times.

Because amino acids and their polymers are ubiquitous on Earth, we have to be reassured that species observed in meteorites are not due to contamination, either while the meteorite lay on the ground, or during subsequent handling. Many meteoritic amino acids are "non-proteinogenic', never having been observed in the biosphere. Others that overlap with biology have a general tendency to be present in a racemic mixture whereas the L-enantiomer dominates in terrestrial proteins. This is a pointer toward extraterrestrial origin, but an "L" excess has also been seen in certain meteoritic amino acids [18,19,20,21,22] making the enantiomer ratio a less-than-perfect indicator that could generate false negatives – i.e. an amino acid could be truly extraterrestrial even though it showed an "L" excess. It appears that the strongest indicator is an isotope signature of enhanced D/H, $^{13}C/^{12}C$, $^{15}N/^{14}N$ or $^{18}O/^{16}O$ – and moreover a signature directly from the chemical in question via a "compound specific" test. Such tests have shown that D, $^{13}C$ and $^{15}N$ are regularly enhanced in meteoritic amino acids ([18,23,24,25]. Measurements of very high isotope enhancements by secondary ion mass spectrometry [26,27,28,29,30] do not identify the chemical compounds that contained the heavy isotopes. In order to link isotope ratios to specific chemicals the usual measurement of isotope ratios involves firstly separation of the chemical under study [18] then pyrolysis followed by mass spectrometry, often requiring gram quantities of meteoritic material. Here, in support of their extraterrestrial provenance we report the signature for isotope enhancement in polymer amide molecules identified via MALDI/TOF mass spectrometry in the mass 1,000 range, using a very small (10mg) initial sample mass. Although this MALDI information is partial, in that all isotopes are additive in MALDI and cannot be distinguished from each other, nevertheless anomalous values well above terrestrial have a clear collective signature.

Here we describe in more detail the composition and properties of Allende and Acfer 086. Our CV3 samples are fragments that do not show any heat-burnished exterior surfaces and are curated in dry, sealed conditions at the Harvard Mineralogical and Geological Museum. Acfer 086 is one of a large collection of meteorites found in the Algerian Sahara, extensively referenced in [31]. It is a CV3 chondrite, unpaired, that shows a low level of shock (category S1) and a moderate degree of weathering (category W3). Wlotzka et al. [32] have obtained via $^{14}C$ dating the terrestrial ages of many Acfer meteorites (not Acfer 086) and a weathering grade W3 was found to correlate with a terrestrial average residence time of 13,000 years. In contrast, Allende [33] was a recorded fall after which samples were rapidly collected. The carbon and nitrogen content of Acfer 086 cannot be found in the literature, although the concentrations of Fe, Ni, Sm, Sc, Na, Mn, Zn, Se and S have been recorded [34,35]. One large difference between Acfer 086 and Allende relates to the sodium content, measured at 543ppm in Acfer 086 vs 3122ppm in Allende [34]. It has been suggested that sodium can be lost during weathering [35]. The levels of many other elements in Acfer 086 and Allende have been given, relative to Orgueil, by Hammond et al [36] and Orgueil levels themselves are given by Anders and Grevesse [37]. Although the abundances in general are extremely similar, Acfer 086 has excesses over Allende for U, Ca, Sr, Ba, Li and K and deficiencies in Ir, Mo, Ni, Co, Ag, Cs and Na. In the analysis below we will refer to the concentrations of alkali ions, which are summarized in Table 1. No reports can be found of amino acid measurements in Acfer 086. Allende, on the other hand, has been very extensively studied for elemental composition and amino acids. Initially [14] Cronin and Moore were unable to detect water-soluble amino-acids in Allende, but Harada and Hare [15] established by hot water extraction and inner drilling that three amino acids persisted at increasing depths from the surface, namely glycine, alanine and an unknown species they tentatively identified as β-alanine. At the surface they measured a broader range of terrestrial amino acids at two orders of magnitude greater concentration than at depths greater than 6mm, indicating possible surface contamination, and providing a model for our own drilling method, described below. In MM2015 the amino acids in Allende were below our detection threshold, but glycine was dominant in the present report. Amino acids in Allende have also been reported by Burton et al. [16].

Table 1. Summary of alkali metal data in Allende and Acfer 086.

| Alkali ion | Allende | Acfer 086 |
|---|---|---|
| Li (ppm) | 1.6 [36,37];  2.5 [38];  1.9 [39] | 8.6 [36,37] |
| Na (ppm) | 3122 [34];  3140 [35]; 3430 [39] | 543   [34] |
| K (ppm) | 328 [36,37];  268 [39] | 792   [36,37] |

We arrived at the study of polymer amide in meteorites by way of earlier work on a simple hydrophobic polymer amide (ATP synthase subunit c) that is (genetically) highly conserved and shows rugged, almost material-like properties. These earlier observations led to speculation as to whether such a simple and rugged class of molecule could have formed at the start of a solar system or even in pre-solar times out of the main elements H/C/N/O in gas phase space (helium excepted). The biological polymer, subunit c of ATP synthase, is the rotor component of the complex of proteins that makes the cell's energy source ATP. We observed that it could function alone in nanometer scale holes in a silicon chip [40] and form a beta-sheet polymer vesicle skin around water that entrapped and ordered the water molecules to <1pm precision [41], henceforth MM2008. A molecule with these rugged, versatile properties seemed to us to be ideal to function, if it could form in gas phase space, as a binding element via its hydrogen bonds, of the stellar dust of silicon, minerals and water. A theoretical study [42] (henceforth MM2014) of the interaction of pairs of amino acids in the gas phase conditions of a warm dense molecular cloud showed that indeed there could be exothermic polymerization, suggesting that polymer amide <u>as a material</u> could in principle have had a pre-solar existence, dating back even to the era of first generation stars when C, N and O were first produced in quantity alongside H and He. It was suggested in MM2014 that such polymers could aid accretion by virtue of their many hydrogen bonds and flexible backbone. Techniques from the isolation of this small hydrophobic protein were applied to meteorite extraction.

# Materials and methods

**Chondritic carbonaceous meteorites as the source of polymer paired with volcano controls**
Charles H Langmuir of Earth and Planetary Sciences, Harvard, advised which CCMs to use for analysis. Every meteorite sample was paired with a terrestrial volcano control that had solidified within seconds at $1220^0C$ and was therefore of atomic composition with intact molecules observed therein being a measure of terrestrial contamination. The curatorial assistant Theresa Smith, of the Harvard Mineralogical and Geological Museum, provided meteorite and volcano control samples in sealed containers delivered on the day of processing to the clean room. The use of each sample is recorded. After use they were photographed to display the etch position on the sample surface, resealed and returned to the Harvard Mineralogical and Geological Museum.

Details of the meteorites and volcano control are as follows:

**Acfer-086 (Fig 1A),** Source: Agemour, Algeria, found 1989-90 TKW 173g, via dealer: David New 17.5g.

**Allende (Fig.1B)**, Source: Mexico 17.30g, carbonaceous (c-chondrite). Fell Feb. 8th 1969.

**Volcano Control (Fig.1C)**, source: MGMH#17364, Basalt, Kilauea Sink, Hawaii. Collected by T.A. Jaggar, January 1921. The additional information is as follow: "Toes and Blisters in glassy basalt. NW. End SW-right flow of 1921 in Kilauea Sink."

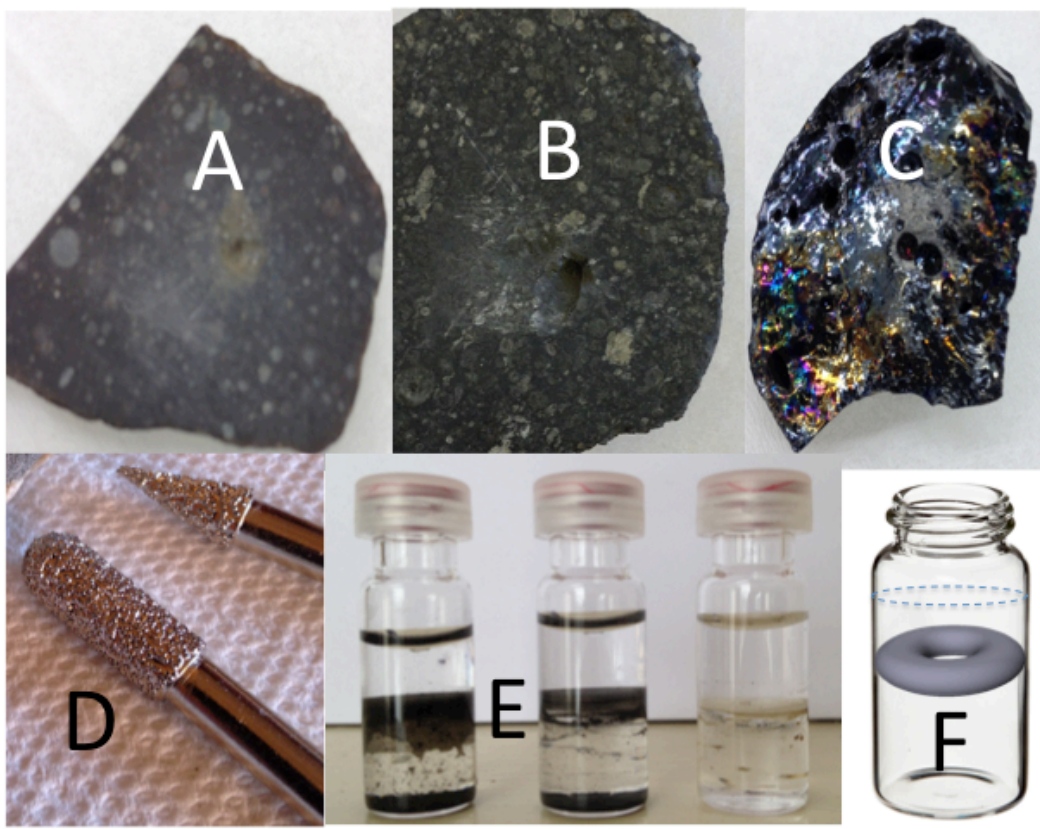

**Fig 1. Acfer 086 (A), Allende (B) meteorite samples together with Hawaii volcano control (C) showing pits etched via diamond burrs (D) rotated via stepper-motor, to produce micron scale fragments. Folch extraction of the fragments showing 2 phases (E) where initially for the meteorite samples (black due to carbonaceous content) the micron fragments reside at the interphase of the phases as a toroid topology - shown diagrammatically (F).**

**Etching of meteorite samples to produce micron scale particles (Fig.1D)**

Operatives were suited and wore powder-free nitrile rubber gloves. In a clean room extractor hood, at room temperature with high airflow, the samples were hand held while being etched to a total depth of 6mm with diamond burrs. The diamonds had been vacuum brazed at high temperature onto the stainless steel burr shafts to avoid the presence of glue of animal origin and organics in general. Etching on a cut/slice face (not an original exterior weathered face) was via slow steady rotation of a burr under light applied force via a miniature stepper motor that did not have motor brushes and did not contribute metal or lubricant contamination to the clean room. Two shapes of burr were used, the larger diameter type, in two stages, to create a pit of diameter 6mm and depth 6mm, and the smaller conical burr to etch a finely powdered sample, of approximately 1µm particles, from the bottom of the pit without contacting the sides. After each stage the pit contents were decanted by inversion and tapping the reverse side and a new burr was used that had been cleaned by ultrasonics in deionized distilled (DI) water followed by rinsing in DI water and air-drying in a clean room hood. The powder from the third etch was decanted with inversion and tapping into a glass vial. Sample weights were in the range 2-8mg.

**Folch extraction of polymer amide from meteorites (Fig.1E & F)**
Solvents chloroform/methanol/water were added to the glass vials containing the etched particles in the ratio 70/20/10 per the Folch extraction protocol [9] to give a meteorite particle/Folch extract solvent concentration of 2-3mg/ml. Chloroform is pipetted first into the glass bottle containing the meteorite particles, followed by methanol and then water to get the phases established. Gentle swirling of the contents results in the meteorite particles locating at the interface of the bottom phase (mainly chloroform) and top phases (methanol and water). As the particles solvate, a torus topology forms at the interphase of the chloroform and methamol/water phases (MM2015). After 6 days extraction in the clean room hood at room temperature the Folch extracts were pipetted off as 100µl aliquots from the top downwards through the phases, each phase being labeled separately and in order. These aliquots were then immediately stored at -20C until taken for mass spectrometry or amino acid analysis and then the aliquots were kept on ice prior to loading onto instruments.

**Matrix Assisted Laser Desorption Mass Spectrometry (MALDI)**
Mass spectrometry was performed on a Bruker Ultraflextreme MALDI-TOF/TOF instrument. As the matrix we used sinapinic acid at 10mg mL$^{-1}$ in water/acetonitrile, 0.1% formic acid, for mass detection 50-50,000amu. Our resolution was of the order of 10,000 to 20,000 and we looked in the range m/z = 50-50,000, finding peaks from 200-24,000. A sample volume of 1µl was mixed with a matrix volume of 1µl and the spots allowed to dry. In some cases the dry spots were rinsed once with ultrapure water so as to remove alkali ions that can generate matrix cluster peaks in the spectrum. In MALDI/TOF there is considerable shot-to-shot variation and slight mass shifts occur based on laser energy. We found experimentally that on different days, after calibration against standard peptide mixtures, the instrument records the same peak with a precision of 0.2Da in the mass range 2,000Da. A precision of 10,000 rules out analysis via more advanced proteomic identification techniques that identify the atomic composition of single peaks via use of a resolution approaching $10^6$. Our response to this limitation is to analyze the whole spectrum at once via an integer mass procedure, discussed in the following section, that is designed to verify the presence of a trial amino acid composition of the polymer. Effectively we are comparing the whole spectrum with a "template" constructed from all the possible masses that can be generated from random combinations of the trial amino acid set.

**Statistical analysis of MALDI results**

For analysis in the range 200 < m/z < 1200 the mass spectrometry peaks are assigned to the nearest integral mass value as described in MM2015. A raw spectrum is filtered to remove peaks due to the sinapinic acid matrix and also peaks present in the volcano control, so as to remove handling or background contamination. The remaining peaks are processed as follows:

Suppose we suspect that the unknown polymer amide contains three "trial" amino acid residues: glycine, hydroxy-glycine and alanine, with residue masses 57, 73 and 71 respectively. The polymer amide species will be $HO(Gly)_i(GlyOH)_j(Ala)_kH$ where the OH and H terminations due to aqueous solution are included. Rounding the mono-isotopic masses to integer masses, this species has mass ($57i + 73j + 71k + 18$). Using a pure poly-glycine standard we have found that more often than not an additional proton is added to form the positive polymer amide ion, so a mass addition of either 18, or 18 + H exists above the sum of bare residue masses. This proton addition is also observed in the present MALDI analyses of meteoritic polymer amide.

For simple trial amino acid combinations there often is a unique set of indices i, j and k that gives a match. As an example, suppose that an experimental m/z peak appears at 318 and it is singly charged. There is only one combination of the above set of trial amino acids that matches this mass: { i= 4, j = 0, k = 1, termination 18 + H}. This method does not deduce the polymer composition directly from the m/z peak, however, because <u>other</u> combinations of amino acids in different sets may also give the same 318 mass. Only when a large number of matches exists within an <u>ensemble</u> of m/z peaks is it possible to apply a statistical test to confirm the likelihood of a chosen polymer composition being present.

When $Q$ matches are observed for $N$ experimental m/z peaks we can estimate from the binomial distribution just how likely it is to have this number of matches. If the probability of achieving $Q$ matches <u>randomly</u> is very low, then it is deduced that the trial set of amino acids is very probably present in the sample. Before applying the binomial test we need to find the probability $P$ of a match occurring randomly. The matches in set $Q$ define a field of indices that we use to calculate all of the <u>possible</u> masses that can be generated throughout this restricted field. Dividing this total of possible masses by the mass range of the data yields the probability $P$ of a random polymer matching an m/z peak composed of the trial set of amino acids, and giving a false positive. In practice, suppose, for example, that this evaluation gives $P = 0.5$, and there are $Q = 18$ matches found in a data set of $N = 21$ m/z peaks. The chance of this occurring randomly is $6 \times 10^{-4}$, the same as the probability of achieving "heads" 18 times in 21 tosses of a coin. It may then be stated that there is a 1600:1 probability that there is a polymer of this composition in the sample. Any single example of a match would have been 50% probable just on a random basis, so that no conclusion could be drawn from one match, but the extension to large $N$ made the detection of a specific polymer type achievable. Other trial sets of amino acids also have to be run against the data and their probabilities evaluated in a similar fashion. The statistical confirmation of one specific trial set has then to be augmented by different types of data such as TOF/TOF fragmentation and amino acid analysis.

**Amino Acid analysis via Acid Hydrolysis and LC/MS for amino acids**

Gloves were worn at all times. From each layer of the Folch extract a 100μL sample was placed in a 60mm glass tube marked with a diamond tipped pen then dried by Speedvac. Glass tubes were placed in a larger vacuum vessel that holds 12 glass tubes maximum. 200μl of 6N HCl was placed in the bottom of the large vacuum vessel using a long pipette tip. The vessel was made devoid of air with alternating purges of vacuum and nitrogen (3X) using the hydrolysis port and oven. The vessel cap was in the closed position after the last purge and the vessel was removed from the hydrolysis port and placed in the oven. The vapor phase hydrolysis took place at 130 C for 1 hour. After 1 hour the vessel was removed from the oven and let cool for a few minutes. The glass tubes were removed from the large vessel with plastic forceps, wiped clean of acid with a Kimwipe and dried in the Speedvac for 10minutes. At this point 50μL of water was added to the samples, followed by vortexing and removal to an Eppendorf.

Following hydrolysis, amino acids were assayed using an Agilent Technologies 6460 Triple Quad machine operated in the multiple reaction monitoring (MRM) mode. The LC column was type Phenomenex Luna SCX (50 x 2.0mm). Solvent A was 30mM ammonium acetate and solvent B was 5% acetic acid. The flow rate was $0.4mLs^{-1}$. Between 0.00 and 6.00min. the flow comprised 87.5% solvent B and 12.5% solvent A. Between 7.00 min. and 11.00min. the flow comprised 100% solvent A. Finally solvent B at 87.5% with solvent A at 12.5% were added to equilibrate for 3min. Amino acids were ionized by electrospray and detected by charge amplifier. In MRM the precursor ion was 19mass units heavier than the bare amino acid residue. After fragmentation, characteristic product ions were followed in real time. As an example, glycine residues of mass 57 were terminated via water solution to mass 75, then protonated to mass 76 in the ionizer of the mass spectrometer. In the quadrupole the dwell time was 50msec, and fragment ions of masses 30.1 and 48.1 were measured. The mass 30.1 signal was used to quantify glycine. Calibration was via 10μMolar pure amino acid samples.

For the above LC/MS assay, 10μL was extracted from the 50μL hydrolyzed sample in the Eppendorf tube. The smallest detectable concentration in the solution loaded for LC/MS was approximately 20nMolar. Considering that this was present in a 10μL volume, the smallest detectable amount of an amino acid loaded onto LC/MS was $2 \times 10^{-13}$ Moles. For a typical molecular weight of 100, this becomes 20picograms. In the results section we report the detected concentrations before and after hydrolysis for the three bottom fractions where 1μMolar (50 times the minimum detectable) is equivalent to 1.0nanograms loaded onto the LC/MS for a typical amino acid. From the meteorite sample masses and the Folch solution volumes, 1μMolar corresponds to 13,560ppb, or alternatively 135nmol/g for Acfer-086; 14,875ppb, or 148nmols/g for Allende; 8,750ppb or 87nmols/g for volcano control. However, these numbers assume equal distribution of extracted polymer amide throughout the whole Folch volume whereas the data shows a concentration of polymer amide around the region at the top of the chloroform layer and the bottom of the water/methanol layer. When the sample is only taken from this layer, an artificially high polymer amide reading is obtained, consequently the quoted nmol/g above can in theory overestimate the meteoritic concentration by up to ten times.

**Method for analysis of heavy isotopes**

With such small samples it is not possible to separate a single polymer composition and subject it to pyrolysis and gas phase mass spectrometry, the classical method to determine compound-specific isotope levels. In our data we observed anomalous $\Delta m = +1$ and $\Delta m = +2$ satellites to m/z peaks that showed higher than terrestrial values relative to the primary peak, hence we introduced the analysis presented below in section 3.4 to test trial assumptions of heavy isotope enhancement vs the data. This technique is most sensitive for masses in the range of 1000Da, for which the terrestrial $\Delta m = +1$ satellite is approximately 50% of the primary peak. Because the heavy isotopes are superimposed in these satellites we can only obtain a collective estimate of their presence, which, however, is a good indicator of extra-terrestrial origin.

**Computational methods for potential polymer structure and its IR spectrum**

All molecular structures were built and analyzed using Spartan software [43]. Molecular geometry was refined prior to any *ab initio* energy calculation via molecular mechanics using the MMFF94 force field [44]. The MMFF94 force field was used to determine energy and the infrared (IR) spectrum when comparing different lengths of polymer.

# Results

## Mass spectrometry and analysis

Mass spectrometry was performed as described in "methods" on samples from the bottom, middle and top phases of the Folch extracts (Fig.1E). The middle and top phases yielded m/z (mass to charge ratio) peaks that were common to the volcano control and the meteorites, whereas the peaks unique to meteorite resided in the uppermost layer of the bottom (chloroform) phase. In the following analyses the polymer peaks were all derived from that layer. Sinapinic acid spectra are often modified by the addition of one matrix mass *m* minus a hydrogen (*m* – H = 223.07) to a true peak [45,46] and such "wings" were removed from the m/z < 2000 data.

**The 4641Da species**

Figure 2 shows mass spectra in the 0 < m/z < 20,000 range of Acfer 086, Allende and the volcano control using sinapinic acid matrix, taken on the same spectrometer run and displayed on the same scale. In addition to the 4641 peaks there are multiples at 9.3kDa; 14kDa; 18.6kDa (only Acfer) and 23.2kDa (only Acfer), and a peak at m/z = 2313 that is assigned to a doubly charged 4.6kDa entity. The region 4000Da < 6000Da from Figure 2 is shown again at higher resolution in Figure 3.

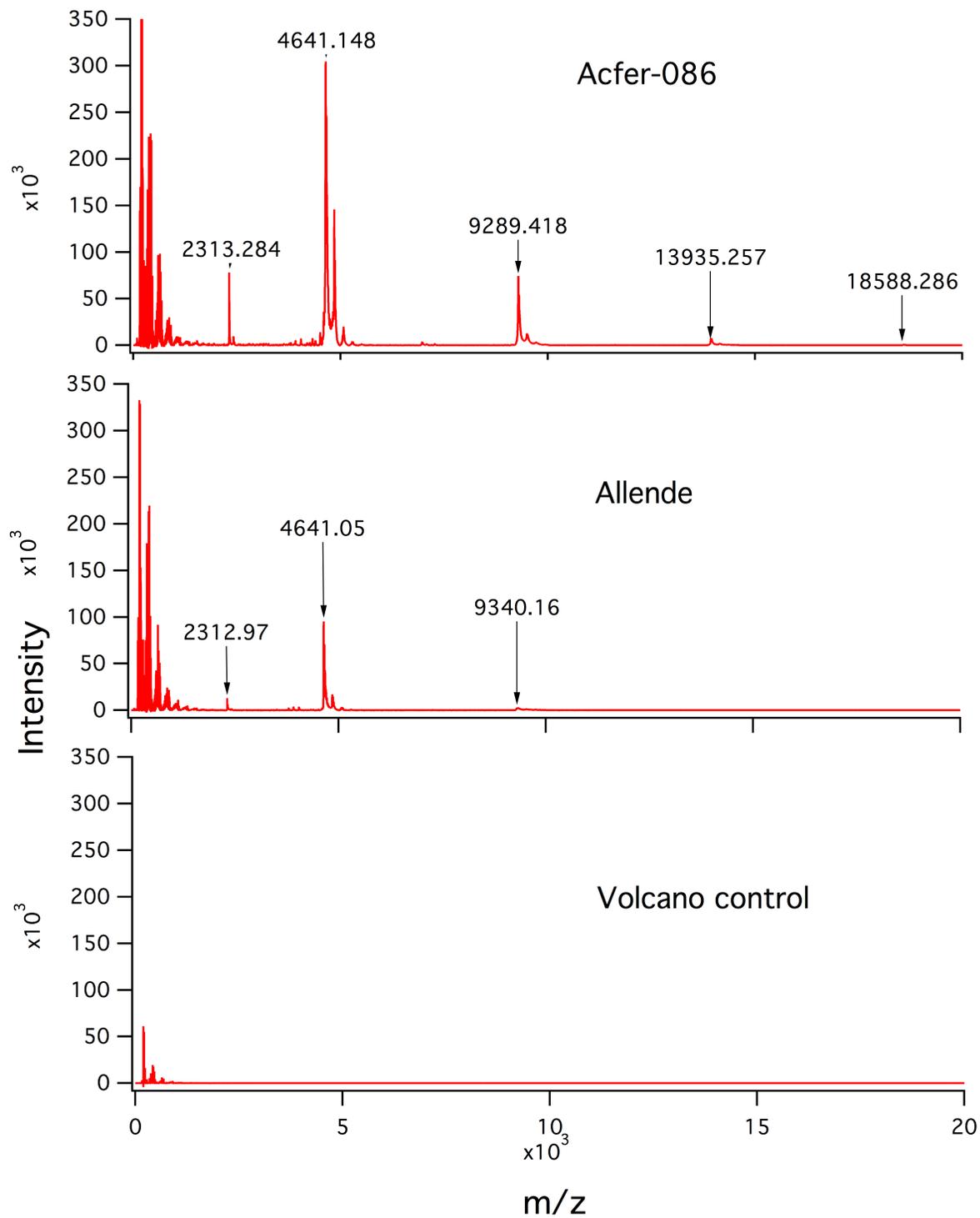

**Fig 2. Mass spectra in the 0 < m/z < 20,000 range of Acfer 086, Allende and volcano control using sinapinic acid matrix. Displayed on the same scale.**

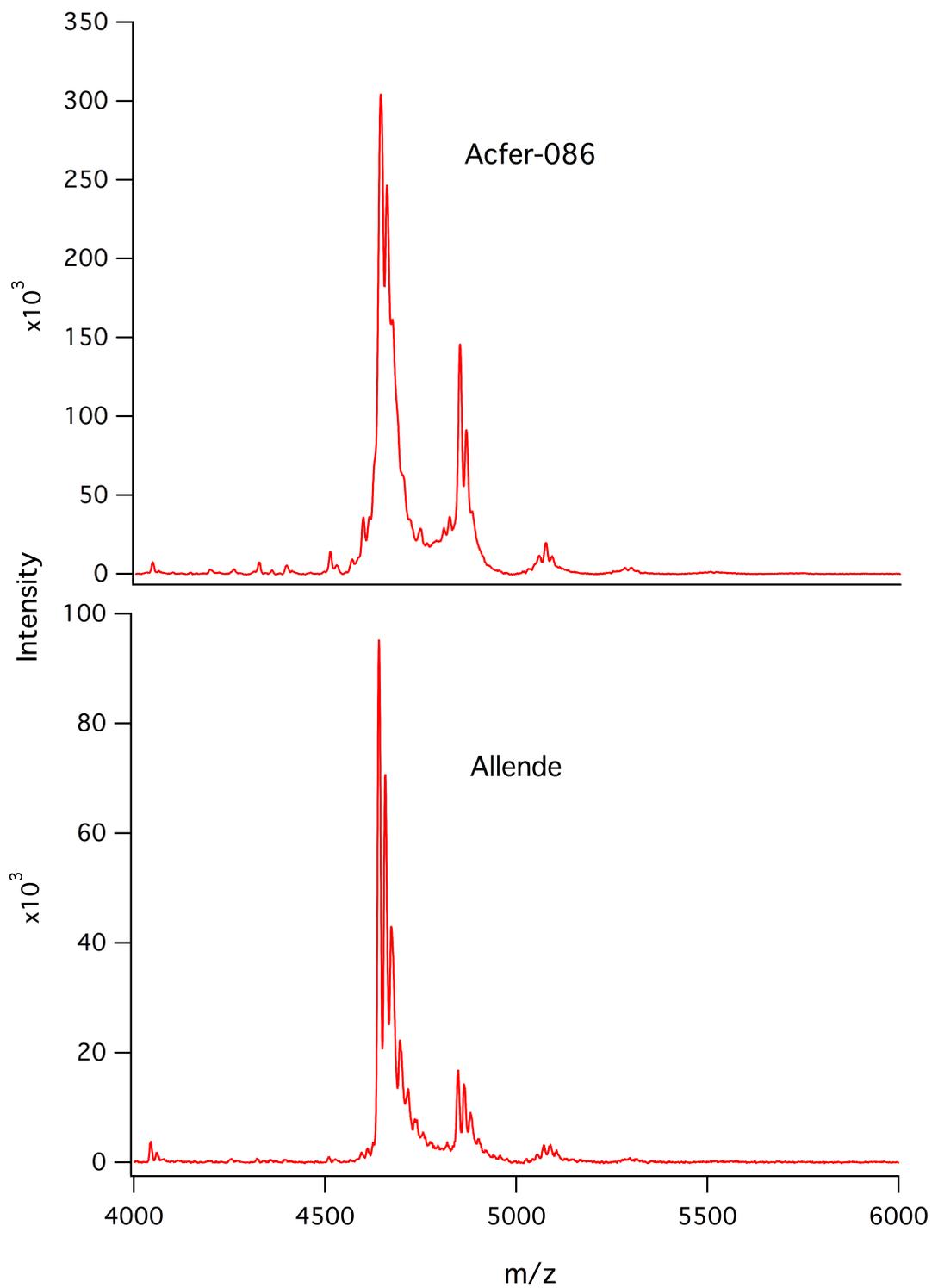

**Figure 3. Mass spectra in the 4000 < m/z < 6000 range of Acfer 086 and Allende using sinapinic acid matrix.**

The m/z > 4000 data is assembled in Tables 2 and 3 for Allende and Acfer 086 respectively, showing the relative amplitude of the peaks and the proposed structural identification. Within the structures there is an entity V* of mass 2297.81 that is inferred below from the analysis of the m/z < 2000 fragments seen at the low end of the data in Figure 2 and listed in Tables 4 and 5. Because of certain coincidences, it cannot be determined from our data whether the various adducts to the 4641 species are composed of hydroxylations shown as 'OH', with a mass addition of 16 together with a lithium addition of mass 7, or equivalently of a sodium atom of mass 23. Although we have given the matches in Tables 2 and 3 in terms of 'OH' and Li, there are other interpretations involving Na = 'OH' + Li, or even K = 2'OH' + Li.

**Table 2. Analysis of the Allende m/z > 2,000 spectrum shown in Figure 2.**

| Structure | Calc. m/z (*) = 2+ | Observed m/z | Intensity | Also in Acfer-086 |
|---|---|---|---|---|
| 2V* + 'OH' + 2Li + H | 2,313.19 (*) | 2,312.972 | 13,025 | # |
| 2V* + 2'OH' + 2Li (2V*+2Na) | 4,641.38 | 4,641.051 | 95,273 | # |
| 2V* + 3'OH' + 2Li | 4,657.38 | 4,657.42 | 70,793 | # |
| - | - | 4,673.822 | 43,016 | |
| - | - | 4,696.092 | 22,394 | |
| - | - | 4,717.127 | 13,468 | # |
| (2V* + 'OH' + 2Li )+(m-H) | 4,848.45 | 4,847.837 | 16,816 | # |
| (2V* + 2'OH' + 2Li)+ (m-H) | 4,864.45 | 4,864.126 | 14,416 | # |
| 2(2V* + 3'OH' + 2Li) + 'OH' + Li | 9,337.70 | 9,340.163 | 2,276 | (#) |
| 3(2V* + 3'OH' + 2Li) + 2'OH' + 2Li | 14,018.02 | 14,017.109 | 179 | (#) |

**In the table m = 224.07 is the sinapinic acid matrix mass and 'OH' represents a hydroxylation with a mass addition of 16. Li isotopic mix assumed with M=6.94. Possible substitution of Na for 'OH' + Li is illustrated for the 4641Da peak.**

**Table 3. Analysis of the Acfer 086 m/z > 2,000 spectrum shown in Figure 2.**

| Structure | Calc. m/z (*)=2+ | Observed m/z | Intensity | Also in Allende |
|---|---|---|---|---|
| 2V* + 'OH' + 2Li + H | 2,313.19 (*) | 2,313.284 | 82,221 | # |
| 2V* | 4,595.50 | 4,595.116 | 49,287 | |
| 2V* + 'OH' | 4,611.50 | 4,611.582 | 49,804 | |
| 2V* + 2'OH' + 2Li | 4,641.38 | 4,641.148 | 317,999 | # |
| 2V* + 3'OH' + 2Li | 4,657.38 | 4,657.215 | 260,378 | # |
| - | - | 4,716.357 | 48,394 | # |
| - | - | 4,744.88 | 42,350 | |
| - | - | 4,805.988 | 42,823 | |
| - | - | 4821.152 | 49,821 | |
| 2V* + 'OH' + 2Li + (m-H) | 4,848.45 | 4,848.209 | 159,422 | # |
| 2V* + 2'OH' + 2Li + (m-H) | 4,864.45 | 4,864.715 | 105,077 | # |
| 2V* + 3'OH' +2Li + (m-H) | 4,880.45 | 4,879.354 | 53,530 | |
| 2(2V* + 2'OH' + 2Li) + Li | 9,289.70 | 9,289.418 | 75,426 | (#) |
| 3(2V* + 2'OH' + 2Li) + 2Li | 13,938.02 | 13,935.257 | 7,687 | (#) |
| 4(2V* + 2'OH' + 2Li) + 3Li | 18,586.34 | 18,591.267 | 794 | |
| 5(2V* + 2'OH' + 2Li) + 4Li | 23,234.66 | 23,228.506 | 124 | |

**In the table m = 224.07 is the sinapinic acid matrix mass and 'OH' represents a hydroxylation with a mass addition of 16. Li isotopic mix assumed with M=6.94. Possible substitution of Na for 'OH' + Li is to be noted.**

Table 4. Unique peaks in Allende for the mass range m/z < 2000.

| Glycine | Alpha Hydroxy glycine | Alanine | Termination Mass | Calculated m/z (*)=2+ (**)=3+ | Observed m/z |
|---|---|---|---|---|---|
| 3 | 2 | 0 | 18 | 167.554 (*) | 167.777/167.666 |
| - | - | - | - | - | 181.546 |
| - | - | - | - | - | 213.160/213.082 |
| 5 | 0 | 1 | 18 | 227.065 (*) | 226.884 |
| 11 | 0 | 1 | 18 | 238.761 (**) | 238.716 |
| 3 | 3 | 1 | 18 + Li | 243.080 (*) | 242.779 |
| 9 | 3 | 0 | 18 + H | 250.420 (**) | 250.738/250.569 |
| 8 | 3 | 1 | 18 | 254.756 (**) | 254.581/254.554 |
| 5 | 3 | 0 | 18 + Li | 264.584 (*) | 264.607/264.448 |
| 4 | 1 | 0 | 18 | 319.113 | 319.092 |
| 3 | 0 | 2 | 18 + H | 332.157 | 332.117/332.018 |
| 14 | 3 | 0 | 18 | 345.120 (**) | 345.132/344.928 |
| 6 | 0 | 0 | 18 | 360.139 | 359.848 |
| 5 | 1 | 0 | 18 | 376.134 | 375.776 |
| 7 | 3 | 2 | 18 | 389.142 | 388.844/389.555 |
| 10 | 1 | 2 | 18 | 401.658 (*) | 401.643 |
| 10 | 3 | 0 | 18 | 403.637 (*) | 403.698 |
| 11 | 3 | 0 | 18 + Li | 435.648 (*) | 435.573/435.555 |
| 5 | 1 | 1 | 18 | 447.171 | 447.447 |
| - | - | - | - | - | 457.456/457.531 |
| 15 | 1 | 0 | 18 + H | 473.678 (*) | 473.404 |
| 4 | 3 | 1 | 18 + H | 537.190 | 536.822 |
| 10 | 2 | 1 | 18 | 805.295 | 805.238 |
| 13 | 2 | 0 | 18 | 905.322 | 905.041 |
| 12 | 3 | 0 | 18 | 921.317 | 921.705 |
| Totals | | | Weighted by number of appearances of a peak | | |
| 235 | 61 | 18 | | | |
| G | $G_{OH}$ | A | | | |
| 100% | 26% | 8% | Relative to Glycine at 100% | | |

**Matrix sinapinic acid. Matches are given for 22 out of 25 peaks in terms of the three chosen amino acids and lithium.**

**Table 5. Unique peaks in the range m/z < 2000 in Acfer 086.**

| Glycine | α-Hydroxy-glycine | Alanine | Termination Mass | Calculated m/z (*) = 2+ (**) = 3+ | Observed m/z |
|---|---|---|---|---|---|
| 3 | 2 | 0 | 18 | 167.554 (*) | 167.719 |
| - | - | - | - | - | 181.49 |
| 5 | 0 | 1 | 18 | 227.065 (*) | 226.931 |
| 11 | 0 | 1 | 18 | 238.761 (**) | 238.792 |
| 9 | 3 | 0 | 18+H | 250.420 (**) | 250.437 |
| 8 | 3 | 1 | 18 | 254.756 (**) | 254.631 |
| 5 | 3 | 0 | 18 + Li | 264.583 (*) | 264.49 |
| 4 | 0 | 1 | 18+H | 318.141 | 318.148/317.919 |
| 4 | 1 | 0 | 18+H | 320.120 | 320.161 |
| 6 | 0 | 0 | 18 | 360.139 | 359.903 |
| 15 | 3 | 1 | 18 + Li | 390.139 (**) | 389.778 |
| 10 | 3 | 0 | 18 | 403.637 (*) | 403.701 |
| 11 | 2 | 1 | 18 + H | 431.662 (*) | 431.649 |
| 11 | 3 | 0 | 18 + Li | 435.648 (*) | 435.61 |
| - | - | - | - | - | 457.642 |
| 3 | 3 | 1 | 18 + Li | 486.161 | 486.044 |
| 4 | 3 | 1 | 18 + Li | 543.182 | 542.961 |
| - | - | - | - | - | 599.414/599.226 |
| 9 | 1 | 0 | 18 + Li | 611.220 | 611.356 |
| 8 | 1 | 1 | 18 + Li | 625.236 | 625.274 |
| 24 | 0 | 0 | 18 + Li | 696.763 (*) | 696.779 |
| 13 | 0 | 0 | 18 | 759.289 | 759.40 |
| 13 | 0 | 0 | 18 + Li | 766.289 | 766.426 |
| 12 | 3 | 0 | 18 | 921.317 | 921.005 |
| Totals | | | Weighted by number of appearances of a peak | | |
| 192 | 34 | 10 | | | |
| G | $G_{OH}$ | A | | | |
| 100% | 18% | 5% | Relative to Glycine at 100% | | |

**Matrix Sinapinic acid. Matches are given for 21 out of 24 peaks in terms of the three chosen amino acids and lithium.**

**The m/z < 2000 fragments**

The m/z peaks unique to meteorite at m/z < 2000, after removal of peaks also present in the volcano control, as well as matrix "wing" peaks, are listed in Tables 4 and 5 for Allende and Acfer 086, respectively. We found that almost all peaks could be matched with combinations of glycine, hydroxy-glycine and alanine, with a uniform aqueous termination of mass 18, or a termination of 18 plus a hydrogen or lithium atom. The mass 18 aqueous termination plus a hydrogen atom is a pattern seen in the spectrum of pure poly-glycine in aqueous solution. Meteoritic polymer amide was first determined (MM2015) to contain these three amino acids and the 18 or 18+H termination via use of the binomial statistical analysis described in the "methods". When, in the present work, lithium appeared in the composition of the 4641Da species, we introduced lithium alongside amino acids as a trial constituent of the fragment m/z < 2000 mass peaks, with the result that more matches were found and a much better statistical identification of the mixture was obtained versus restriction to hydrogen.

In order to evaluate the probability of achieving so many matches in the Allende case of Table 4, we counted the number of possible masses that satisfied the criterion {arbitrary Gly ; 0->3 $Gly_{OH}$ ; 0->1 Ala ; 0->1 Li ; 0->1H, and termination 18} to find that the probability of fitting a random mass by chance was 42% in the range of the data. Using the binomial distribution we found that the probability of randomly having 19/25 matches in Allende (not counting 2Ala in 3 cases) was $6 \times 10^{-4}$. Applying the same analysis to Acfer 086 fragments, listed in Table 5, we found 21 matches to 24 peaks with a probability of $5 \times 10^{-6}$ that they could have occurred randomly. Each meteorite gave similar ratios of glycine: hydroxy-glycine: alanine as shown in the Tables. There was a marked excess of $Gly_{OH}$ triplets in this data, giving additional weight to the observation of exactly $3Gly_{OH}$ as a mass component in TOF/TOF fragments reported previously (MM2015) and further discussed below.

There is an overall correlation between m/z peaks in Allende and Acfer 086, best illustrated by the "stick" graph of Figure 4. Not only is there a family resemblance in the 4641 region, but many of the fragments also are in common. We use this, together with the very similar $Gly/Gly_{OH}/Ala$ ratios (Tables 4 and 5) to permit the merging of the two data sets for the purpose of finding the most appropriate model, discussed in the following section.

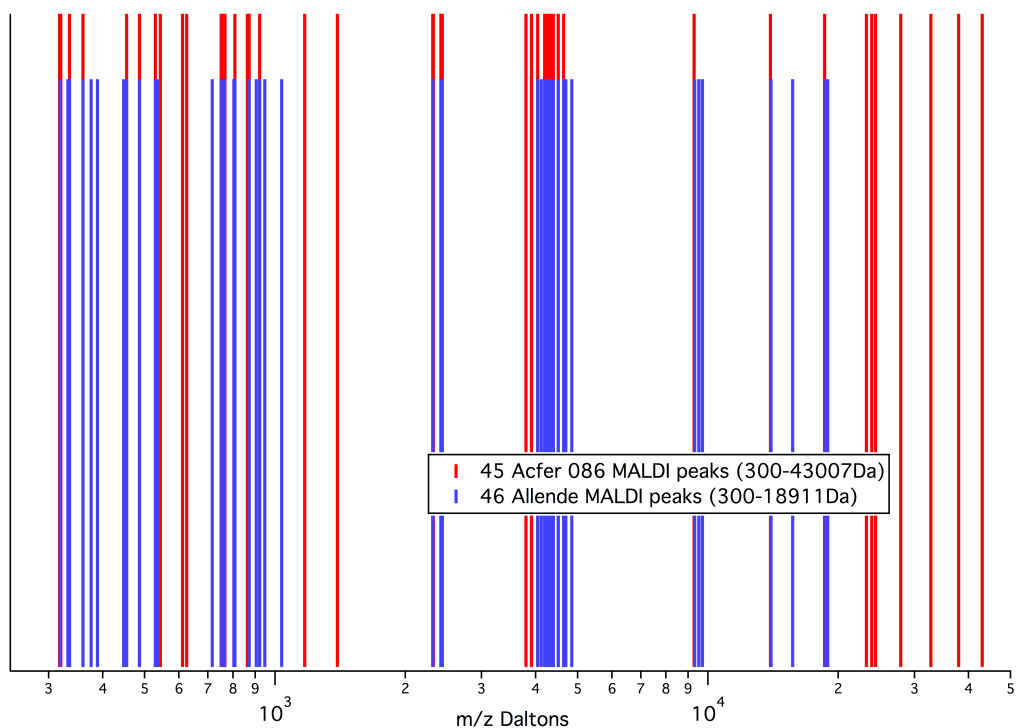

Acfer 086 & Allende observed m/z distributions
The main peak is at 4641Da with degradation products at lower masses and multiples of 4641 at higher masses
Height arbitrary

**Fig 4. Superposition of m/z peaks for Allende (blue) and Acfer 086 (red)**
**The main peak is at 4641Da with degradation products at lower masses and 4641 multiples. Height arbitrary.**

**Derivation of a subunit structure for the 4641Da species from the fragment data.**
We have followed the fitted distribution of Gly, $Gly_{OH}$ and Ala in order to create a "parent molecule" that also meshes with the 4641 set of peaks. In Figure 5 we propose a subunit labeled **V*** that specifically incorporates two groups of three contiguous $Gly_{OH}$ residues that are postulated to be pre-existing within the meteorite, and not merely formed post-extraction in the alkaline solution. If the modification of glycine to $Gly_{OH}$ had been occurring post-extraction, then it is unlikely that we should have seen exactly three hydroxylations across so many experiments. In order to envision the relative distribution of glycine and hydroxy-glycine we have compiled in Figure 6 a visual display of the data from five experiments including the Allende data from MM2015 and both Allende and Acfer 086 data from the present set. The eye is drawn to a horizontal trend for exactly 3 $Gly_{OH}$ masses in the fits to data. The average count of alanine is only 0.7 in this data compilation. Our hypothetical V* has to fragment into the distributions we observe.

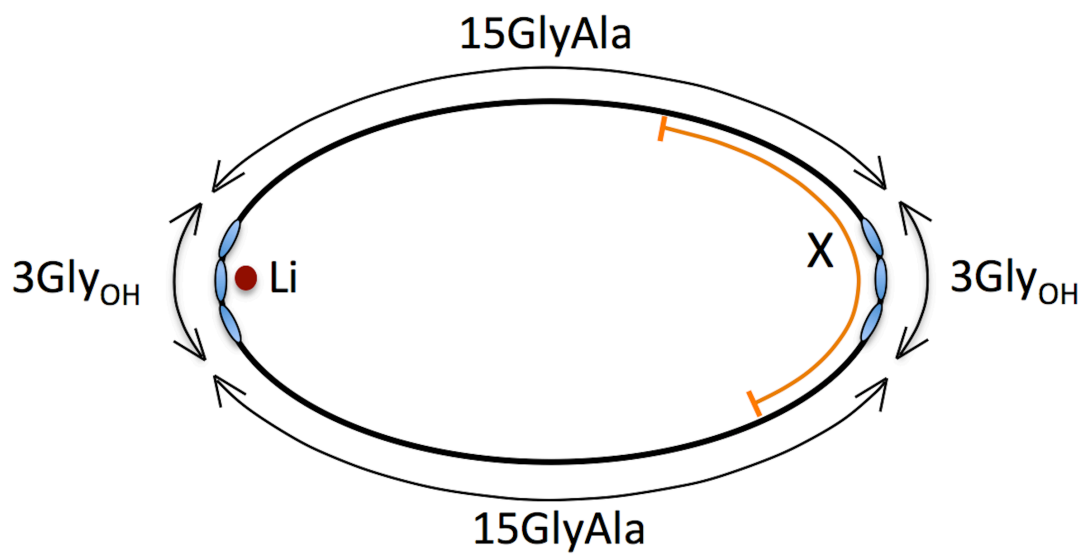

**Fig 5. Structure of V* subunit showing a typical fragment of length X residues.**

Key:    blank = 0    o = 1    • = 2    ● = 3    ⬤ = 4 occurrences

Note: Alanine averages 0.7 in m/z fits

Number of Glycine residues

|  | | 1 | 2 | 3 | 4 | 5 | 6 | 7 | 8 | 9 | 10 | 11 | 12 | 13 | 14 | 15 | 16 | 17 | 18 | 19 | 20 |
|---|---|---|---|---|---|---|---|---|---|---|---|---|---|---|---|---|---|---|---|---|---|
| Gly$_{OH}$ | 0 | • | o | o | • | • | • |   | o |   | • | • |   | • |   |   |   |   | o | o |   |
|  | 1 |   |   |   | o | • |   |   | o | o | o | o |   | o |   | o | • |   |   |   | o |
|  | 2 |   |   | ● |   | o |   | o |   |   | o | o |   | o | • | o | o |   |   | o |   |
|  | 3 | o | • | ● | • | • | o | o | • | ● | ⬤ | ● | • | • | ● | • |   | • | o |   | o |
|  | 4 |   | o |   | o |   |   |   |   |   | • |   | o |   | o |   |   | o |   |   |   |
|  | 5 | o |   |   |   |   |   |   | o | • |   | o |   | • |   |   | o |   |   |   |   |
|  | 6 |   |   |   | o |   |   |   | o |   |   | o |   |   |   | o |   |   |   |   |   |

**Fig 6. Cumulative compositional data of meteoritic polymer amide.**

**The 3Gly$_{OH}$ groupings and their locations in V***
The weighted ratio of Gly$_{OH}$ : (Gly+Ala) in the merged data of Tables 4 and 5 is 0.209, suggesting that for each stretch of 16Gly, that may contain one Ala via methylation, there are 16 x 0.209 = 3.3 Gly$_{OH}$ residues. In the construction of V* this is rounded to 3. There is information on the grouping of these residues to be obtained from their frequent appearance as exactly 3Gly$_{OH}$ in the fragments. Because the composition is evidently similar, we were able to merge the data from Tables 4 and 5 to obtain better statistics. In 43 fits of fragments under the merged fragment analysis, 32 had Gly$_{OH}$ present, but none had more than three Gly$_{OH}$ residues. The Allende and Acfer 086 merged data in Table 6 shows that 3Gly$_{OH}$ occurs much more often than 2Gly$_{OH}$ or Gly$_{OH}$, and somewhat more often than 0Gly$_{OH}$.

**Table 6. Number of appearances of the 3Gly$_{OH}$ grouping.**

| Number of Gly$_{OH}$ | Allende fits | Acfer 086 fits | Merged | x=12 | x=11 | x=10 | x=9 |
|---|---|---|---|---|---|---|---|
| >3 | 0 | 0 | 0 | 0 | 0 | 0 | 0 |
| 3 | 10 | 9 | 19 | 20 | 18 | 16 | 14 |
| 2 | 3 | 2 | 5 | 4 | 4 | 4 | 4 |
| 1 | 5 | 3 | 8 | 4 | 4 | 4 | 4 |
| 0 | 4 | 7 | 11 | 10 | 12 | 14 | 16 |

Observed fits from Tables 4 and 5, and results of a fragmentation simulation with varying residue count X.

The last four columns of Table 6 show the results of a fragmentation simulation in which a fragment of length X = 12, 11, 10 or 9 residues is cut out of V* from any of the 38 possible locations, as illustrated in Figure 5. By chance the simulation numbers, which each add up to 38, are on the same scale as the data numbers of fits, so it is easy to compare them, to find that an 11-residue excerpt gives a good match to the merged Gly$_{OH}$ distribution. This agrees well with the measured (weighted) average number of residues in a fragment, which is 10.5, indicating the self-consistency of proposing that there are two separate groups of 3Gly$_{OH}$ that are equally spaced in the V* loop (but not proving it definitely).

**Lithium atom on one 3Gly$_{OH}$ group**
Analysis of the lithium content of the fragments yields a ratio of Li : (Gly+Ala) equal to 0.0308, using the merged data from Tables 4 and 5 and weighting the content of each species by the number of appearances of a peak. In V* the Ala arises via methylation of Gly in either of the two 16-Gly stretches, so the average number of Li atoms in V* is 32x0.0308 = 0.986. It is therefore accurate to include only one Li atom within V*. As to its location, we expect it to be attracted to the 3Gly$_{OH}$ regions, and note that within those there are 10 cases with 3Gly$_{OH}$ + Li in the fit, and 15cases with 3Gly$_{OH}$ + 0Li in the fit. Apart from these there are only 2 cases involving Li and not Gly$_{OH}$. It is very likely from this distribution that one Li atom is associated with one of the 3Gly$_{OH}$ groups, and that the other group does not carry Li.

**Methylations (Glycine occasionally converted to Alanine)**

Merging all the m/z < 2000 fits in Tables 4 and 5 we find that the fraction of Gly that has been methylated to Ala is Ala/(Gly + Ala) = 0.0615. In V*, with 32Gly, this would imply 1.97 methylations. From this we round off to 2 methylations, but there is no clear correlation with $Gly_{OH}$ to show whether these are positioned other than randomly. Further data may show where they tend to be located within the Gly stretches. In the models of Figure 7 we have applied one methylation to a random position within each 16 Gly stretch. The compact methyl groups do not perturb the anti-parallel beta sheet structure.

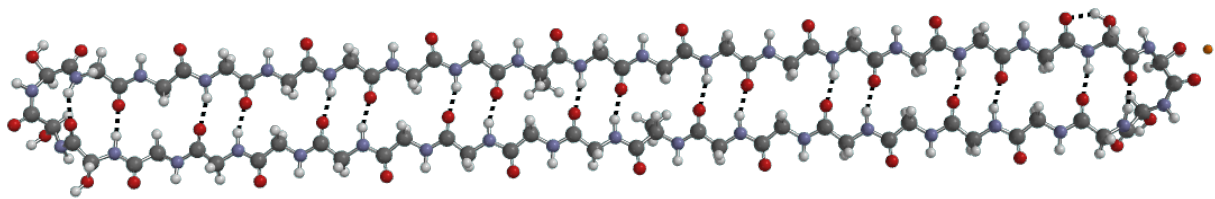

**Fig 7a. Single antiparallel strand of V* in "ball and spoke" format. Atom labels: hydrogen grey/white, lithium orange, carbon black, nitrogen blue, oxygen red, Hydrogen bonds dotted black.**

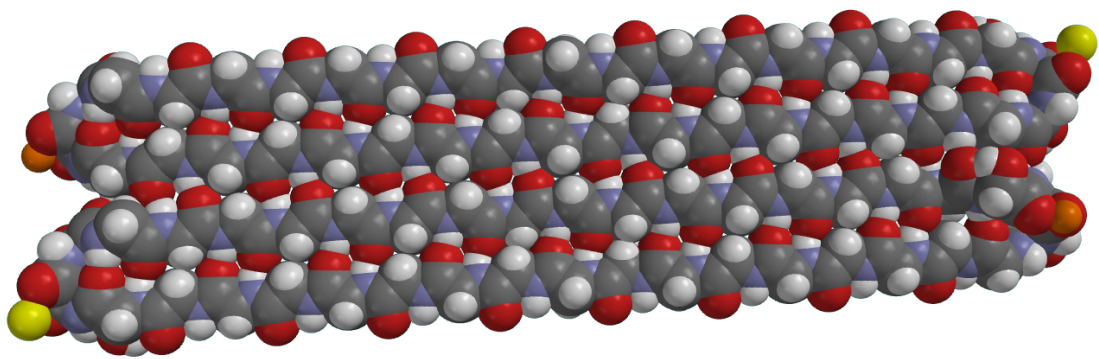

**Fig 7b. Space-filling model of (2V* + 2Na) dimer with mass 4641Da. Atom labels: hydrogen grey/white, lithium orange, sodium yellow, carbon black, nitrogen blue, oxygen red.**

**Structure and mass of V* and its dimers**

The proposed V* molecule is drawn in Figure 7a&b as a loop, without direct proof from the data, because such a structure without reactive termini has a definite mass, sufficient to explain the precise appearance of the 4641 species in two different meteorites. Retrospectively, V* would not have provided the matches in Tables 2 and 3 if that polymer had an aqueous termination with a mass addition of 18. Such a polymer amide loop buttons up as a stable anti-parallel beta sheet, as illustrated in Figure 7a. Two units of V* may also lock together via an anti-parallel set of hydrogen bonds to form a very stable dimer with two lithium atoms present, as illustrated in Figure 7b. We have arbitrarily drawn this dimer object with 6Gly$_{OH}$ residues in proximity at each end, contributing a hydrophilic character to its ends and possibly encouraging further additions of water or alkali atoms, to generate all of the mass peaks in the "4600" range.

Employing mono-isotopic masses we find M(V*) = 2,297.81 and its dimer M(2V*) = 4,595.62. However, data collection is spread over several mass units around a peak in the m/z > 4000 range, so we must employ the average isotopic mass of Li, M = 6.94, rather than the mono-isotopic mass of 7. The 2V* dimer is then at 4,595.50Da which matches the lowest mass peak within the "4600" group of Acfer 086 (observed mass 4,595.12), within the MALDI observational accuracy of approximately one part in 10,000.

The 2V* structures need an additional mass of 46 units in order to constitute the dominant 4641 peak seen in both Acfer 086 and Allende. One possibility is that this could be contributed by two sodium atoms, one at each end of the 2V* dimer (Figure 7b). One of the differences between the Allende and Acfer 086 "4600" complexes is that Allende starts exactly at the main peak 4641 whereas Acfer 086 starts at 4595. If sodium atoms are the means for conversion from 4595 to 4641, then the more complete conversion in Allende would be consistent with the six times higher sodium content of Allende relative to Acfer 086. In Tables 2 and 3 the assignment of sodium is not unique as its mass of 23 could also be composed of one hydroxylation (an alpha hydrogen on glycine replaced by OH with a mass addition of 16), plus a lithium addition, as discussed above.

The higher multiples of the 4.6kDa species seen in Tables 2 and 3 are possibly bound together by alkali ions, a process also seen in the clustering of MALDI matrix molecules in the presence of alkali ions [47].

**Amino acid analysis**

The top layer of bottom fractions of Allende, Acfer 086 and volcano extracts were analyzed for amino acids via acid hydrolysis and LC/MS in an instrument calibrated to detect the 20 biological amino acids at concentrations down to about 20nMolar (Table 7). Before hydrolysis no amino acids were detectable in Allende, Acfer 086 or volcano, with the exception of Glu and Lys in volcano. Essentially no additional amino acids were found in volcano after hydrolysis, and the volcano readings provide a reference control for the meteorite data. The amino acid analyses showed higher overall concentrations in Acfer 086 than in Allende after hydrolysis. It is emphasized that our assay is set up to detect only the biological amino acids, so the presence of many others known to be extra-terrestrial was not investigated. In our previous experiments (MM2015) the Allende amino acid readings had been too low to register but in the present experiment there were readings above background (the volcano data is a background measure of laboratory contamination) for Ser, Gly, Pro, Ala, Phe and Leu, with Gly dominant. This data lends support to the dominant Gly interpretation for polymer amide in Allende.

In Acfer 086 there are significant amounts of amino acids with Glu, Ser and Gly dominant. This caused us to check possible fits to the Acfer 086 polymer spectrum using combinations of these three residues, but no significant grouping emerged. The amino acids other than glycine, while not free, are possibly in random polymers that do not occur with sufficient frequency to show up in the m/z > 2000 range.

**Table 7. Acid hydrolysis of Acfer 086/Allende meteorites and volcano control.**

| Amino Acid | Acfer 086 Non-hyd. | Acfer 086 Hyd. 130C | Allende Non-hyd. | Allende Hyd. 130C | Volcano Non-hyd. | Volcano Hyd. 130C |
|---|---|---|---|---|---|---|
| Asp | 0 | 11.86 | 0 | 0.93 | 0.45 | 0.69 |
| Cys | 0 | 0 | 0 | 0 | 0 | 0 |
| Glu | 0.02 | 19.51 | 0 | 1.26 | 1.35 | 1.49 |
| Ser | 0 | 24.05 | 0 | 1.43 | 0 | 0.63 |
| Asn | 0 | 0 | 0 | 0 | 0.02 | 0 |
| Thr | 0 | 11.48 | 0 | 0.33 | 0.02 | 0.14 |
| Gln | 0.02 | 0 | 0 | 0 | 0.04 | 0.04 |
| Tyr | 0 | 5.96 | 0 | 0.2 | 0 | 0 |
| Gly | 0 | 19.10 | 0 | 2.20 | 0 | 0.92 |
| Pro | 0.05 | 10.14 | 0.04 | 0.28 | 0.01 | 0 |
| Ala | 0 | 9.19 | 0 | 0.61 | 0 | 0 |
| Met | 0 | 0.73 | 0 | 0 | 0 | 0 |
| Val | 0.04 | 6.21 | 0.11 | 0 | 0 | 0 |
| Phe | 0.04 | 3.34 | 0.10 | 0.3 | 0.01 | 0.11 |
| Leu | 0.08 | 6.97 | 0.19 | 0.5 | 0 | 0.18 |
| Trp | 0 | 0 | 0 | 0 | 0 | 0 |
| Lys | 0.13 | 4.38 | 0.12 | 0.41 | 0.48 | 0.3 |
| His | 0.06 | 1.53 | 0.05 | 0.12 | 0.16 | 0 |
| Arg | 0.07 | 10.56 | 0.07 | 0.31 | 0.38 | 0.21 |

**Biological amino acids only. Concentration of amino acids in hydrolysate in µMolar. A zero reading means <20nMolar. Approximately, 1µM equals 100nmol g$^{-1}$ of meteorite, or less, subject to comments in text.**

**Heavy isotope excess from MALDI mass spectrometry satellites**
Examination of the mass spectrum satellites at $\Delta m = +1$ and $\Delta m = +2$ in relation to a fitted polymer amide peak showed intensity enhancements that indicated a heavy isotope excess. The present section derives a framework to analyze this effect.

The following terrestrial standards ([48] Vienna) are taken as reference values:
VSMOW water  $R_H = {}^2H/{}^1H = 155.76 \pm 0.05 \times 10^{-6}$
VSMOW water  $R_O = {}^{18}O/{}^{16}O = 2,005.20 \pm 0.45 \times 10^{-6}$
V-PDB  $R_C = {}^{13}C/{}^{12}C = 11,237.2 \times 10^{-6}$
Atmospheric Nitrogen  $R_N = {}^{15}N/{}^{14}N = 3,612 \pm 7 \times 10^{-6}$

In a molecule with *n* hydrogen ($^1H + {}^2H$) atoms in its formula we calculate the probability of obtaining *k* $^2H$ atoms from the binomial distribution [49] as

$$P_H(k) = \frac{n!}{k!(n-k)!} p_H^k (1-p_H)^{n-k}$$

where $p_H = \frac{[^2H]}{[^1H]+[^2H]}$ is the probability of a heavy isotope substitution. For example, if the ratio of heavy to light isotopes is $R_H = {}^2H/{}^1H$ then $p_H = R_H/(1+R_H)$ = 1.5574 x 10$^{-4}$ for the terrestrial hydrogen standard. Similar considerations lead to values for $p_C$, $p_N$ and $p_O$, denoted in general by $p_X$ in Table 8. In the polymer amide mass spectra (MM2015), certain peaks are seen to be matrix clusters because a) they belong to the sequence of clusters identified by Harris et al. [47] and b) they also appear in the volcano control spectrum, which is nominally devoid of polymer amide. A large amplitude matrix cluster peak occurs at 861 mass units (Figure 8) (detail from MM2015, Fig. 2). This has isotopic (Δm = +1 and Δm = +2) satellites of amplitude 47% and 17% relative to the main peak. The '861' cluster has composition ($M_4KNa_3 - 3H$) where the matrix is *M = HCCA* (α-cyano-4 hydroxycinnamic acid) of mono-isotopic molecular weight 189.043 and formula $M = C_{10}H_7NO_3$.

**Table 8. Isotope satellite calculation for the '861' matrix cluster peak**

|  | $H_{25}$ | $C_{40}$ | $N_4$ | $O_{12}$ | $K_1$ | $Na_3$ |
|---|---|---|---|---|---|---|
| $p_X$ | 1.5574 x 10$^{-4}$ | 0.011112 | 3.599 x 10$^{-3}$ | 2.0012 x 10$^{-3}$ | 0.06731 | 0 |
| $P_X(0)$ | 0.996 | 0.640 | 0.986 | 0.976 | 0.933 | 1 |
| $P_X(1)$ | 0.0039 | 0.288 | 0.0036 | 0.023 (Δm = +2) | 0.067 (Δm = +2 | 0 |
| $P_X(2)$ | 7 x 10$^{-6}$ | 0.063 | - | - | - | - |
| $P_X(3)$ | - | 0.009 | - | - | - | - |

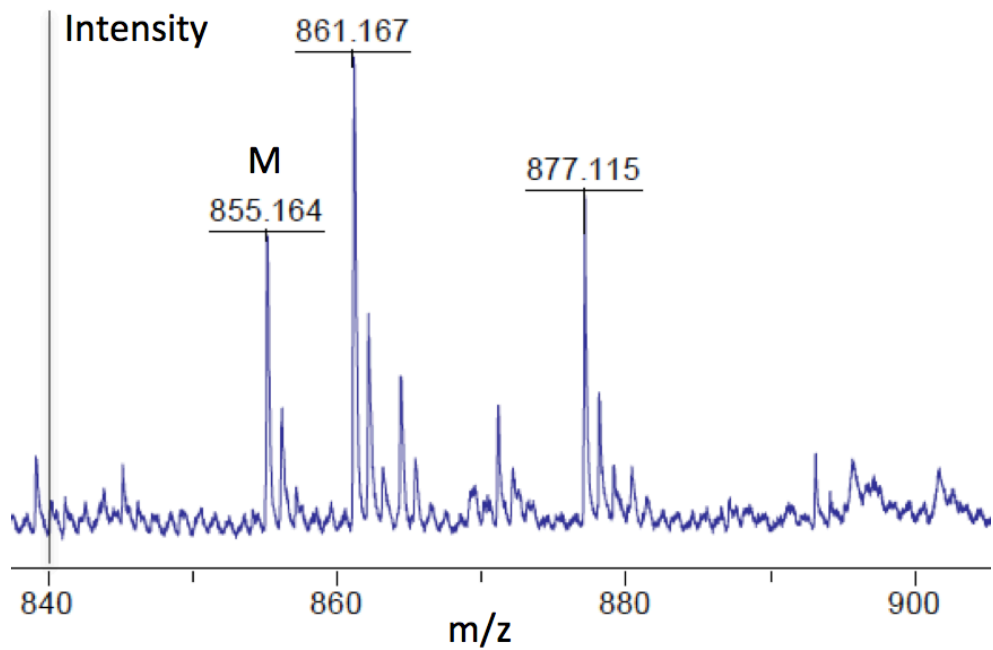

**Fig 8. Matrix cluster peaks at m/z = 855, 861 and 877** reproduced from MM2015, Fig. 2.

Here we calculate the '861' satellite amplitudes in detail to illustrate the procedure and compare with experiment. The composition of the '861' cluster is $H_{25}C_{40}N_4O_{12}K_1Na_3$. Of these constituents only $^{23}Na$ is naturally mono-isotopic. In the following table its heavy isotope probability is zero. Entries in the table are obtained from the binomial distribution [D] for $k$ = 0,1,2,3 heavy substitutions in the $n$ atoms of the species concerned, with $n=25$ for hydrogen, etc.

The isotope ratios are presumed uncorrelated, so the probability of having zero mass addition is the product of the individual $P_X(0)$ terms:
$\Pi(0) = P_H(0)P_C(0)P_N(0)P_O(0)P_K(0)P_{Na}(0) = 0.572$
There are three different ways to have $\Delta m$ = +1, involving $H$, $C$ and $N$ respectively. Oxygen and potassium jump to $\Delta m$ = +2 immediately. The probability of having $\Delta m$ = +1 is therefore:
$\Sigma \Pi(1) = P_H(1)P_C(0)P_N(0)P_O(0)P_K(0)P_{Na}(0)$
$\quad\quad\quad + P_H(0)P_C(1)P_N(0)P_O(0)P_K(0)P_{Na}(0)$
$\quad\quad\quad + P_H(0)P_C(0)P_N(1)P_O(0)P_K(0)P_{Na}(0) \quad = \quad 0.262$
Similarly, the possible combinations that give $\Delta m$ = +2 sum to give $\Sigma \Pi(2)$ = 0.113 and the $\Delta m$ = +3 sum is $\Sigma \Pi(3)$ = 0.033. By now, we have accounted for 0.98 of the collective isotope amplitude, with the remaining 0.02 assigned to $\Delta m$ = +4 and higher species. The total of probabilities should sum to unity, when correctly carried to completion.

For comparison with experiment we normalize the $\Delta m$ = 0 peak to be 100%, as listed in Table 9. For satellites of this (terrestrial) 861 matrix peak there is good agreement with the $\Delta m = +1$ and $\Delta m = +2$ measured intensities but the $\Delta m = +3$ location appears to also carry another peak that obscures the true reading.

**Table 9. Calculated and experimental isotope satellite ratios for the 861 matrix cluster peak.**

|  | $\Delta m = 0$ | $\Delta m = +1$ | $\Delta m = +2$ | $\Delta m = +3$ |
|---|---|---|---|---|
| Calc. | 100% | 45.8% | 19.8% | 5.8% |
| Expt. | 100% | 47% | 17% | - |

**Analysis of the polymer amide isotope satellite data**

The identification of polymer amide in MM2015 was made on a statistical basis via consideration of the whole ensemble of non-matrix and non-volcano peaks. It was not certain that any given peak by itself was polymer amide. However, the likely polymer amide peaks in the m/z range 897 to 1106 shown in Figure 9, reproduced from MM2015, Fig 3 had the enhanced isotopic satellites, averaging 100:74:54, listed in Table 10. Using terrestrial values an isotope calculation was performed for the 1106 polymer amide peak 14Gly3Gly$_{OH}$1Ala(t=18) in which there is a mass 18 aqueous termination t = (OH + H) and the formula is $H_{58}C_{37}N_{18}O_{22}$. The calculated ratio for the terrestrial version was 100:49:16, indicating a strong degree of satellite enhancement in the meteorite data.

**Table 10. Measured amplitude of isotopic satellites.**

| m/z | $\Delta m = 0$ | $\Delta m = +1$ | $\Delta m = +2$ |
|---|---|---|---|
| 897 | 100 | 56 | 29 |
| 1034 | 100 | 84 | 70 |
| 1050 | 100 | 65 | 37 |
| 1066 | 100 | 67 | 26 |
| 1090 | 100 | 92 | 79 |
| 1106 | 100 | 80 | 73 |
| Average | 100 | 74 ($\sigma$ = 12.3) | 54 ($\sigma$ = 20.6) |

**Fitted polymer amide MALDI peaks from MM2015. Integer m/z value of peak given. Satellites normalized as a percentage of the $\Delta m = 0$ amplitude.**

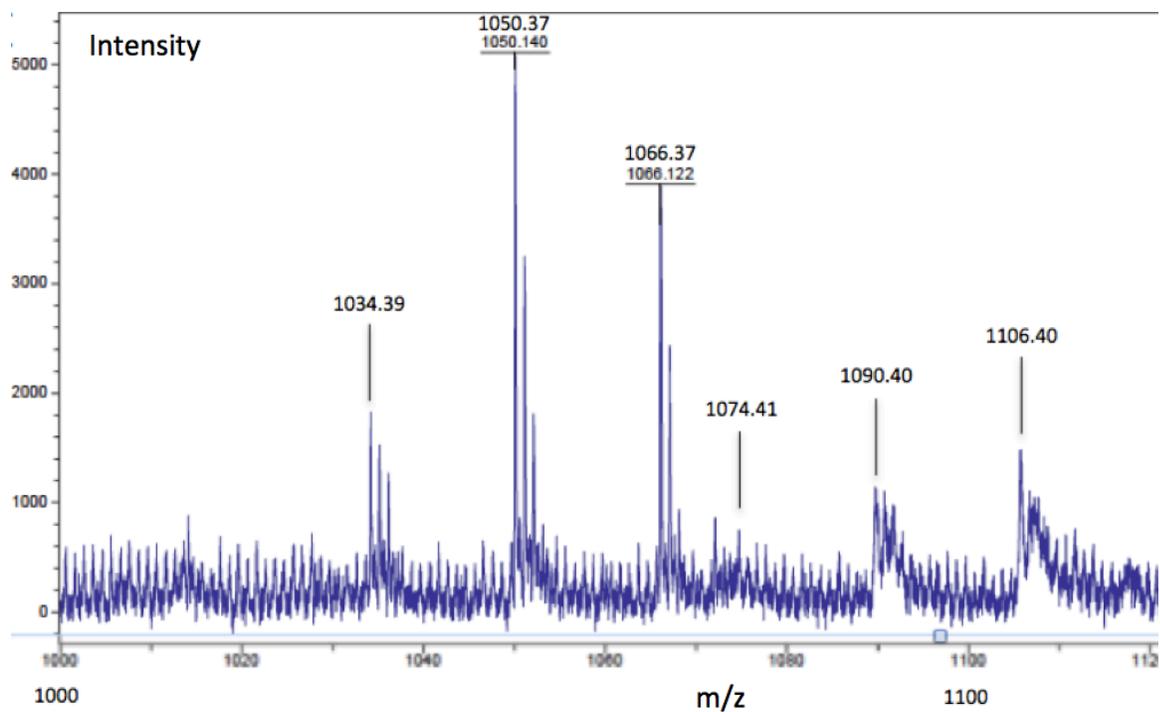

**Fig 9. Identified polymer amide peaks in Allende (from MM2015, Fig. 3).**

In an additional sample preparation of Allende via the above methods, but now with sinapinic acid matrix and extraction at -20C over a period of 42days, a new set of heavy isotope data was obtained. In order to more clearly resolve the isotope peaks the mass spectrometer was run in the "reflectron" mode, giving a resolution of about 50ppm. In this sample we observed dominant polymer fragments at 1108.6Da and 1124.6Da that are identified in view of the discussion in MM2015 as $(14Gly,4Gly_{OH})t=18$ and $(13Gly,5Gly_{OH})t=18$ respectively, with calculated mono-isotopic masses of 1108.38Da and 1124.37Da. Figure 10 shows a typical mass spectrum detail from this new data set. The quality of this data is improved over the 2015 data in both signal-to-noise ratio and resolution. We determined that the average 1108Da satellite ratio was 100 : 68.5 : 29.7 from eight measurements, and the average 1124Da satellite ratio was 100 : 61.3 : 28.0 from 10 measurements (Table 11). The grand average of the two peaks was 100 : 64.5 : 28.7 that reduced statistically to the result 100 : 65±3 : 29±4 as the most probable distribution to represent the data. The data of Table 11 was compared via the t-test to the calculated 1106 terrestrial ratio of 100 : 49 : 16 with the finding that both the $\Delta m = 1$ and $\Delta m = 2$ measured ratios were less than $1 \times 10^{-4}$ probable as samples of the terrestrial ratios.

**Table 11. Measured amplitude of isotopic satellites for polymer amide MALDI peaks from high resolution data, normalized as a percentage of Δm = 0.**

| m/z | Δm = 0 | Δm = +1 | Δm = +2 |
|---|---|---|---|
| 1108 | 100 | 63.1 | 27.0 |
|  | 100 | 63.7 | 27.6 |
|  | 100 | 60.6 | 23.6 |
|  | 100 | 52.2 | 27.8 |
|  | 100 | 74.9 | 39.8 |
|  | 100 | 75.0 | 39.6 |
|  | 100 | 78.2 | 25.7 |
|  | 100 | 79.9 | 26.3 |
| 1108 average (n=8) | 100 | 68.5 ($\sigma$ = 9.3) | 29.7 ($\sigma$ = 5.9) |
| 1124 | 100 | 77.8 | 37.7 |
|  | 100 | 52.5 | 32.4 |
|  | 100 | 53.6 | 29.1 |
|  | 100 | 56.9 | 26.5 |
|  | 100 | 51.8 | 32.4 |
|  | 100 | 72.7 | 28.5 |
|  | 100 | 72.1 | 27.9 |
|  | 100 | 63.9 | 15.0 |
|  | 100 | 60.0 | 27.3 |
|  | 100 | 51.8 | 22.9 |
| 1124 average (n=10) | 100 | 61.3 ($\sigma$ = 9.3) | 28.0 ($\sigma$ = 5.8) |
| Grand average (n=18) | 100 | 64.5 ($\sigma$ = 9.9) | 28.7 ($\sigma$ = 5.9) |

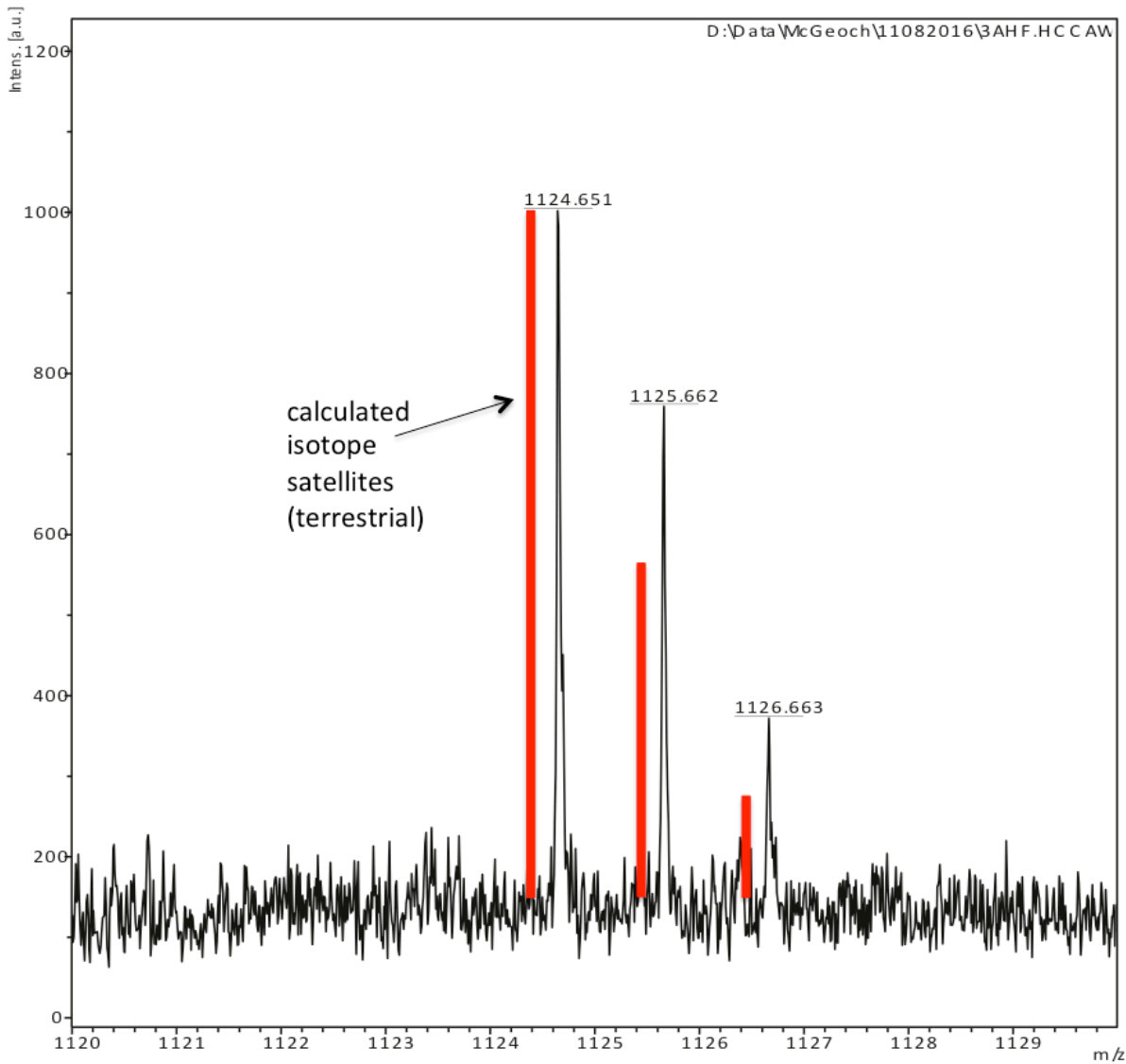

**Fig 10. Enhanced isotope satellites in polymer amide at mass 1124, 1125 and 1126. Calculated terrestrial intensities for this polymer are shown as red bars, offset for clarity.**

Calculated isotope satellites for the '1106' polymer amide peak (which also are very close to expectations for the 1108 and 1124Da peaks) are given in Table 12 for terrestrial, and various combinations of $^2H$ and $^{15}N$ enhancements. We restricted the simulation to $^2H$ or $^{15}N$ ratios because in meteoritic material the observed heavy isotope enhancements relative to terrestrial tend to be much higher for $^2H$ and $^{15}N$ than for $^{13}C$ or $^{18}O$. Also, there have been reports of correlations between $^2H$ and $^{15}N$ enhancements [50,26]. In Table 12 we show that the data matches extreme enhancements in either of $^2H$ or $^{15}N$ separately, and combinations of moderate enhancements involving both isotopes.

**Table 12. Isotope satellite calculations for the 1106 polymer amide peak under various $^2H$ and $^{15}N$ enhancements over terrestrial.**

| Enhancement | | Δm = 0 | Δm = +1 | Δm = +2 |
|---|---|---|---|---|
| $^2H$ | $^{15}N$ | | | |
| 0‰ (terrestrial) | | 100 | 49 | 16 |
| 10,000‰ | 0 | 100 | 58 | 21 |
| 20,000‰ | 0 | 100 | 67 | 24 |
| 0 | 1,500‰ | 100 | 59 | 22 |
| 0 | 3,000‰ | 100 | 68 | 28 |
| 10,000‰ | 1,500‰ | 100 | 68 | 26 |
| 20,000‰ | 3,000‰ | 100 | 87 | 42 |
| Average of observations (Table 10) | | 100 | 74 ± 5 | 54 ± 8 |
| Average of observations (Table 11) | | 100 | 65 ± 3 | 29 ± 4 |

To conclude this analysis, the enhanced isotope satellites in 8 separate mass peaks in the 897-1124 range, obtained with two different MALDI matrix molecules, are indicative of substantially raised heavy isotope levels in meteoritic polymer amide. This approach does not, however, identify the specific isotopes, but gives a global enhancement picture that strongly suggests an extra-terrestrial origin for the polymer.

**Modeling of polymer amide structure**
Construction of possible molecular forms of the 4641 molecule was achieved via Spartan software [43] with every "trial" structure subject to molecular mechanical calculations and *ab initio* calculations if the atom number was <200. The structures displayed in Figure 6a&b represent trials based on our present incomplete knowledge which only further analyses will convert to actual verified structures. Modeling verifies stability of a structure, for example the antiparallel beta sheet of Figure 7a, and allows prediction of observables such as the infrared absorption spectrum. The calculated infrared normal modes revealed a growing mode intensity up to 20 residue length, but a collapse of mode coherence beyond this length, which was one of the considerations going into the choice of V*, to be further discussed below.

**Modeling of the polymer amide infrared spectrum**
The V* structure in Figures 5 and 7a with a mass of 2298Da was preceded in our modeling by simpler alternative labeled U* = 2(19Gly-$\beta$Ala) with β-alanine at the loop turns:

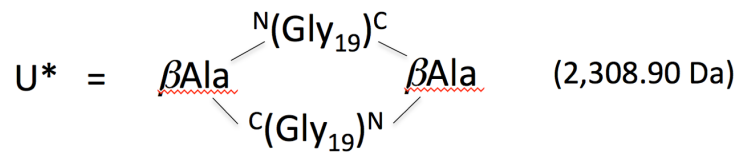

U* is an alternative backbone model for a loop of polymer amide of mass approximately 2300 but U*, not containing Gly$_{OH}$, did not match the observed lower mass degradation products whereas V* did. The U* structure was used for the IR simulations where only the backbone of the molecule is important to the calculations, and variation of the number of residues was studied. This proposed U* of dimensions 7.3 x 0.75 x 0.17nm has hydrogen bonds all along its length between the carbonyl and amide groups of the opposed glycine residues in an 'anti-parallel beta sheet" structure. Figure 11 shows the model for a canonical 40-residue U* and its uncorrected infrared absorption spectrum calculated in MMFF using Spartan software.

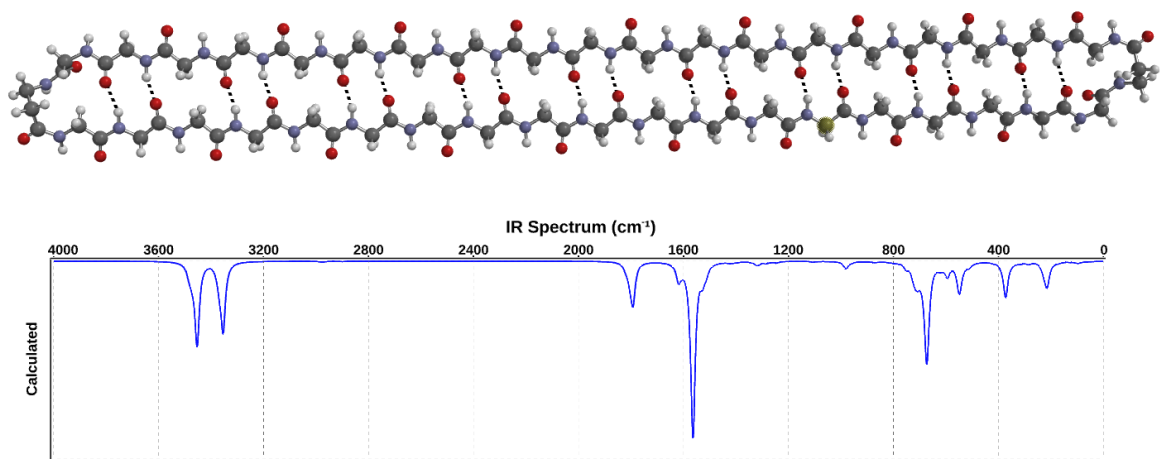

**Fig 11. Molecule U* and its calculated infrared absorption spectrum using MMFF. The calculation has not been normalized to data and gives wavenumbers 5 -10% too high. Correction factors are given in the text. Ball and spoke Spartan view for molecule. Atom labels: hydrogen grey/white, carbon black, nitrogen blue, oxygen red, Hydrogen bonds dotted black.**

In simulations of the IR spectrum of U*, we found that the most intense vibrational modes were collective ones, involving coupling along the molecular chain. The basic structure of the infrared spectrum is found to comprise bands at 3.1, 3.2, 6.2 and 7.1 microns associated respectively with the unperturbed N-H mode, a perturbed N-H mode, the amide I mode and an N-H rocking mode. Figure 12 shows the intensity of the strongest normal mode within each of the bands, as a function of the number of residues in one strand of U*. The intensity of the strongest mode increases dramatically up to a residue count of 20, but collapses in a rather random fashion above this size. The reason for this collapse is the onset above about 20 residues of a "kink" in the MMFF equilibrium geometry of U*. Below 20 residues the molecule is straight and there is collective re-enforcement of a given type of vibration as the result of coupling with the same type of vibration in neighboring residues. When N residues radiate (or absorb) collectively, the radiated intensity scales as $N^2$, and such scaling appears in Figure 12 for N up to 20. The onset of a kink in the molecule breaks it into two smaller "good" regions that each radiate with a greatly reduced mode intensity that depends on where the kink has formed. In every case the strongest mode always belongs to the N-H rocking movement in which most of the N-H groups rock in synchronism toward first one end, then the other end of the molecule. This band has an absorption at 7.1µm and could be the most important one for energy collection. Although energy collection continues to increase with residue count for each of the bands the usefulness of this energy could depend on how much of it is concentrated into a single phonon mode, and if energy collection is a factor in the abundance of the 4641 molecule this data could be a good reason for it not to have more than 20 residues in its subunits. A comparison of 6 different polymer lengths, 10, 15, 19, 22, 25 and 30 is given in Figure 13 showing the kink in an otherwise straight polymer loop that occurs at lengths greater than about 20 residues, which is also the length we propose for V*. It is known that weak nonlinear interactions between the vibrations of neighboring polymer subunits can lead to the development of a kink [51]. In the present case the MMFF potentials are nonlinear and the mode calculation responds to this.

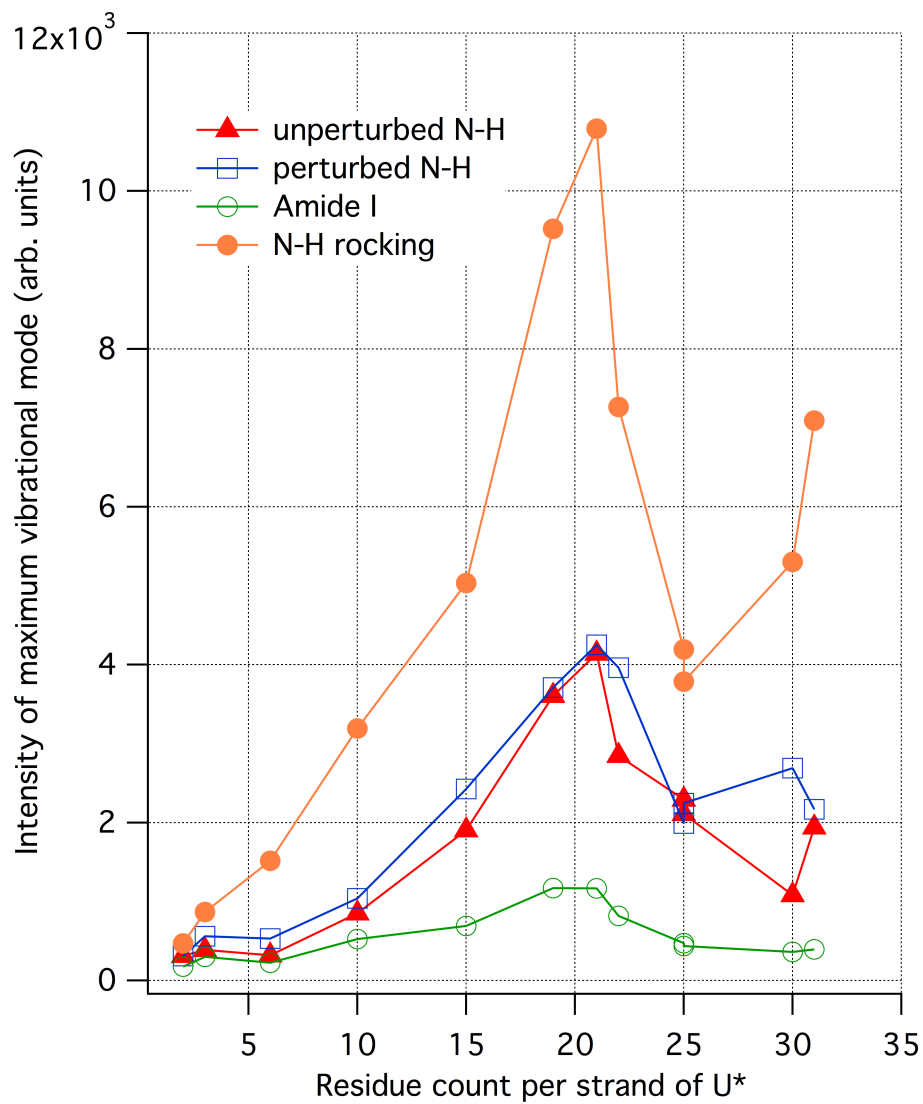

**Fig 12. Intensity of strongest individual vibrational mode for varying numbers of residues in a strand of molecule U*.**

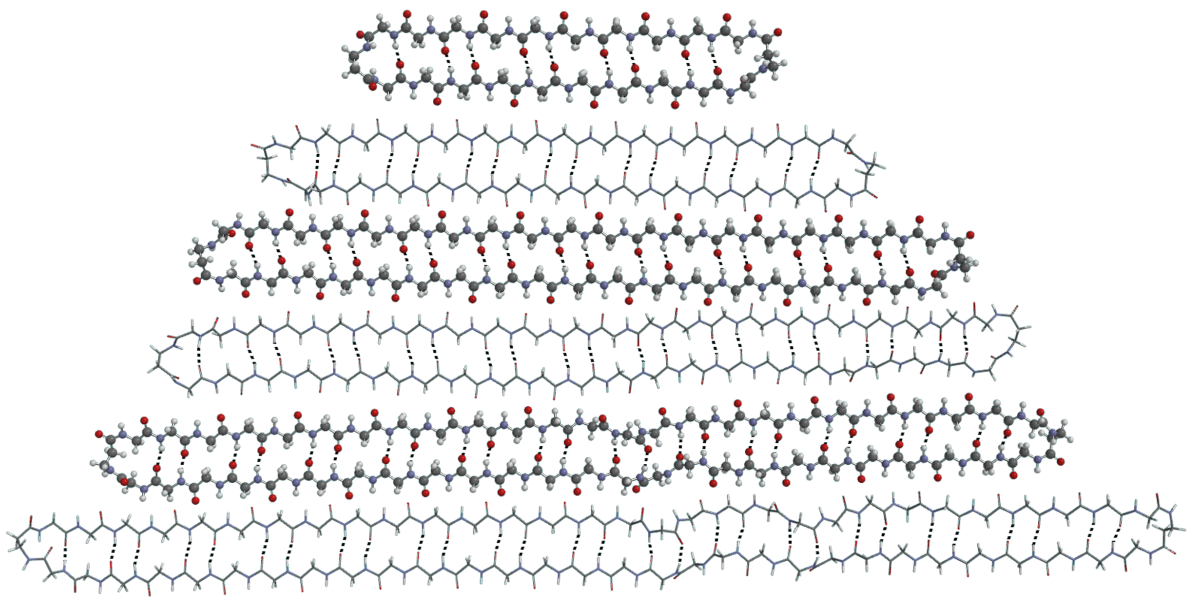

**Figure 13. Polymers based on model U* are from top to bottom: 10, 15, 19, 22, 25 & 30mers in a strand. At the length greater than the 19-21mer the straight polymer kinks. The polymer is displayed alternately as Spartan "ball & spoke" and "tube" for clarity. Atom labels: hydrogen grey/white, carbon black, nitrogen blue, oxygen red, Hydrogen bonds dotted black**

## Discussion

The polymer amide that we have characterized in this work is very probably of extra-terrestrial origin, for the following reasons:
1. It was isolated via drilling a depth of 6mm into meteorite samples and discarding shallower material
2. A control of volcanic rock, curated in the same conditions and subject to the same drilling and extraction did not show polymer amide at all
3. Typical polymer amide fragments in the 1000 mass range showed a heavy isotope enhancement in mass spectrum satellites that was significantly in excess of terrestrial levels
4. The amino acid composition of the polymer was not able to be matched in a search for biological sequences and it contained rare hydroxylation of multiple glycine residues
5. At least one of the meteorites (Allende) was a relatively recent fall that was collected shortly after its appearance.

This confirms that one is indeed looking at extra-terrestrial material, but it does not settle the question of its ultimate provenance, whether from more than 5 billion years ago in the interstellar medium (ISM) or from the proto-planetary disc (PPD) at the time of the solar system's formation, approximately 4567My ago [11]. Heavy isotope enhancements were originally found in small molecules within the interstellar medium (reviewed in [50]) but conditions in the proto-planetary disc could also have caused the concentration of heavy isotopes, particularly D, into small molecules ([52,53,54,55], so a determination of enhancement by itself does not prove an ISM origin as opposed to one in the PPD.

Conditions in the ISM and PPD are harsh, both in terms of radical chemistry and ionization caused by cosmic rays. Although polymer amide molecules are tough (MM2008), it seems that the survival of polymer amide might depend upon its accretion into small "blobs" associated with water (MM2008 and MM2014) and other less abundant species such as silicon and iron. Such "nano-globules" would give protection to inner contents from ultraviolet radiation although they would not protect from direct impacts of cosmic rays. In a similar fashion to cosmic "dust" they could support icy surface "mantles", giving further protection. If they formed in the ISM and survived passage through the PPD, they could be incorporated into primitive meteorites of the CV class alongside dust relics. There have in fact been reports of nano-globules with raised $^{15}$N content in meteorites, but their chemical composition is not yet known [26,27,28,29,30]. These nano-globule isotope measurements were made by nano-SIMS, which has 30-100nm spatial resolution, but only ejects small (two or three atom) ionic fragments, insufficient to identify a polymer. This technique may later help to localize polymer in cases where enhanced isotopes have been identified in polymer amide.

The most intriguing observation is that a single polymer type dominates the polymer amide spectrum, with fragments at m/z < 2000 originating from a proposed parent molecule centered at 4641. Apart from the 4641 complex and its multimers, there are no additional strong peaks in the m/z > 2000 region. When the presence of several other amino acids at comparable level to glycine is considered (data in Table 7) it is curious that the polymer spectrum is so unitary. Out of the millions of polymer amide masses that could be constructed from such a range of amino acids, why should a single mass dominate the others by a factor of much more than one hundred? This observation leads inevitably to the thought that the 4641 molecule has been produced in quantity by "replication" or has otherwise been concentrated. By "replication" we imply a process in which one molecule of polymer amide acts as a template, either directly or indirectly, for the condensation of a second identical molecule out of the glycine that is available. Thermodynamically, replication requires a certain energy input [56] so as to effect separation of the copy from the original template. It is this requirement that may give a clue as to the length of this polymer amide entity. Infrared light is absorbed with rapidly increasing efficiency up to the polymer length that begins to kink, discussed above in section 4.1, which is about 20 amino acid residues. This, we believe, could be the fundamental reason for the size of the present (proposed) replicating entity – it is determined by the need to collect energy at optimum efficiency so as to replicate most rapidly.

Partly guided by this consideration, and more precisely guided by the fragment distribution, we have proposed that there could be a subunit, named V*, of total 40 residue size (in a loop of 20 residue length, when stabilized by hydrogen bonds) and mass 2298Da, that combines to form the 4641 and higher mass polymers via the compositions assigned in Tables 2 and 3. At this stage, we emphasize that V* is simply our best current proposal that fits the data, and that additional analysis has to be done when more data is available. In these tables we have presented the "lithium interpretation" because lithium emerged as an essential component in fitting the fragments (Tables 4 and 5). There is a larger quota of 4641 in Acfer 086 than Allende and the reported native lithium ratios in Table 1 reflect this. If polymer amide originated in the ISM after the first large increase in C, N and O had taken place (MM2014) then its reactivity with the lithium produced in the primordial nucleosynthesis [57] could have led to the lithium deficiency noted in astronomical observations [58].

## Conclusions

In two CV3 carbonaceous chondrites of very different history, but mostly identical elemental composition, we have observed similar polymer amide fragments in the m/z < 2000 region, and the same isolated mass peak at m/z = 4641 with associated higher clusters. A volcanic rock control sample subjected to the same extraction and analysis procedures did not contain polymer amide. A CV3 provenance dates the metoritic polymer to within a few My of the formation of our solar system, and most likely implies an earlier origin in the ISM. Also it implies that the polymer amide could have survived without alteration. By integrating the fragment information we have constructed a trial polymer amide subunit V* that satisfactorily fits the complex of peaks around 4641Da and is designed to break up into precisely the observed distribution of fragments. The dominance of a single species of polymer amide suggests that possibly it formed universally before our solar epoch via replication or that it was otherwise concentrated in interstellar molecular clouds.

# Acknowledgements

We thank Guido Guidotti for helpful discussions and support for the mass spectrometry; Sunia Trauger for some of the mass spectrometry data collection; Charles Langmuir for advice on meteorite selection and provision of clean room facilities; Zhongxing Chen for assistance with the extraction process; and the Harvard Mineralogical and Geological Museum, who provided the meteorite and volcano samples. This research did not receive any specific grant from funding agencies in the public, commercial, or not-for-profit sectors.